\newcommand{\beq}{\begin{quote}}
\newcommand{\enq}{\end{quote}}
\newcommand{\be}{\begin{equation}}
\newcommand{\en}{\end{equation}}

\newcommand{\om}{\omega}
\newcommand{\Om}{\Omega}

\documentstyle[12pt,epsfig]{article}
\begin{document}
\title{ Period relation for the 2:1 resonance in the
GJ876 planetary system 
} 
\date{}
\author{Michael Nauenberg\\
Department of Physics\\
University of California, Santa Cruz, CA 95064 
}
\maketitle
\begin{abstract}
The recent radial velocity Keck data for GJ876 in
Laughlin et al. (2005) is shown to be in good agreement,
apart from a 6 m/s scatter, with a theoretical calculation
(Nauenberg, 2002) based on orbital parameters from
a fit to the earlier Keck data. The time variation of the periods of the inner
and outer planets, which are locked in a near 2:1 resonance,
are evaluated, and their mean values, $P_i$ and $P_o$, are shown to
satisfy closely the relation $P_o/P_i=2+P_o/P$, where $P$ is
the period for the retrograde precesion of the commom mean periastron of 
each of these planets.

\end{abstract}

Recently, Laughlin et al. (2005) have given a progress report
on the extrasolar planetary system GJ976, which consists of
two Jupiter size planets in a near 2:1 resonance, and provided
new high precision radial velocities obtained with the Keck
telescope during the period 2001 to 2004. This new report 
discusses  a fit to a coplanar configuration
of the two planets which includes 
a $6$ meters/second stellar jitter. The possible occurrence of such
a jitter was  suggested in 
(Nauenberg 2002) because of the excesively large values 
of the reduced chi-square in the fit to the data, 
\footnote { In earlier fits to the data 
( Marcy et al. 2001; Laughlin \& Chamber 2001; Rivera \& Lissauer 2001), 
the reduced chi square was reported incorrectly}, and it 
appears  to have further confirmation by studies of
chromospheric activity in M3-M4  stars similar to GJ876
(Laughlin et al. 2005). 
There also may be, however, other sources, e.g. additional smaller planets of
order $6/200$ of a Jupiter mass 
which can also  contribute to the estimated $4$ to
$6$ m/s variations in the maximum radial velocity 
of about $200$ m/s \footnote{After the completion of this
paper, a new planet  of 7.5 earth-mass orbiting GJ876
has been reported (Rivera et al., to be published)}.
In Figs. 1-3 we  show that the
radial velocities in the new Keck data are in good agreement, apart
from a mean 6 m/s  scatter,
with the values predicted from theoretical calculations
(Nauenberg 2002) based on the orbital parameters obtained 
from a fit to the early Keck data of Marcy et al. (2001)
including an additional 9 data points released previously.
The scatter  is shown in Fig. 4.
This good  agreement is  surprising in view of the fact
that in the new report (Laughlin et al. 2005)
the velocities from the earlier data have been
modified.  This modification includes a substantial shift
in the  mean velocity of the central star of about 70 meters/second  
which is included in comparing  the new data with 
our predictions.

It has been customary to characterize
the orbital motion of the two planets of GJ876  
by the  parameters of the
osculating ellipse at a given epoch. In particular,
the two  parameteres $P_i$ and $P_o$ associated
with the periods of the inner and outher planets
are approximately  $30$ and $60$ days, respectively,
which was the original basis for the assertion that
these two planets move in a near 2:1 resonance.
But the actual periods, defined as the time for a sidereal
revolution of the planets, are not constants, but exhibit
oscillations\footnote{ The oscillations of the period, defined 
as the time elapsed  between consecutive periastrons of
the orbit,  is shown  (Nauenberg 2002) }. The inner planet moves 
alternatively between two
nearby orbits, Fig. 5, with two different oscillating  periods
which are  shown in Fig. 6 .
The oscillations of the period of the  outer planet is somewhat
more regular and is shown in Fig. 7. These oscillations have
a period of about 660 days.  Of special interest is  
the mean value of the orbital  periods,
because the condition of a near 2:1 resonance
implies the relation 
\be
\label{ratio1}
\frac{P_o}{P_i}=2+\frac{P_o}{P} 
\en
where $P$ is the mean period for the rotation of the periastron
of these planets, which is moving  retrograde relative to the 
rotation of the planets .  Since this relation, which
reveals the reason  why $P_o/P_i >2$,  has not been presented previously, 
a derivation is given in the following Appendix.  We have calculated the mean
values of these periods by counting the number of rotations
for a given time, and increasing this time $t$ 
until there are no further changes in the period  to a chosen accuracy. 
Increasing $t$ from $1000$ to  $8000$ days, we obtain  
$P_o=61.035$ days, $P_i=30.220$ days and $P=3125$ days. Hence, 
we obtain $P_o/P_i-2=.0197$ and $P_o/P=.0195$, which is in very
good agreement with our period relation, Eq. \ref{ratio1}.
Previously we had estimated that $P=3200$ days (Nauenberg 2002),while
Lee and Peale (2002) obtained $P=3100$ days, and  more recently Laughlin
et al. (2005) reported $P=3205$ days (quoted as a rotation
rate of $41^o$ per year). Actually, we have evaluated $P$ from the orbital
parameters given in (Laughlin 2005),  and obtained instead $P=3136$ 
in better agreement with our results.
  
This  analysis indicates that GJ876  offers a remarkable  opportunity
for a detail quantitative study of a near 2:1 resonance 
in a planetary system, which will greatly improve as more data becomes
available in the future.

\subsection*{Appendix, Period relation for a near  $r:s$ resonance}
In this appendix we consider the general condition that two planets 
are in a near $r:s$ resonance, where $r$ and $s$ are two integers.
Let $\om_i(t)$ and $\om_o(t)$ be the time dependent angular
frequencies for the inner and outer planets, and
$\Om_i(t)$ and $\Om_o(t)$ the corresponding angular frequencies
for the rotation of the periastron of these two planets.
Then the number of rotations  $n_i$ and $n_o$ of the inner and
outer planet  from  one  periastron \footnote
{ The periastron is the position on the orbit nearest to the
central star. It is assumed that this point
is connected to the farthest point on the orbit
by a line, referred to as the line of apsides
(Lee 2000; Laughlin 2005). However,  when the mean
periastron is rotating, as in GJ876, these
two points are found to be $2^o$ to $3^o$ from such
an imaginary line} to the the next one,
during a given time $t$, is given by
\be
\label{wi}
n_i=(\om_i-\Om_i)t/2\pi,
\en
and
\be
\label{wo}
n_o=(\om_o-\Om_o)t/2\pi
\en
where $\om_i$, $\om_o$, $\Om_i$ and $\Om_o$ are the mean values of these
angular frequencies during $t$,
\be
\om_i=\frac{1}{t}\int_o^t dt' \om_i(t'),
\en
\be
\om_o=\frac{1}{t}\int_o^t dt' \om_o(t'),
\en
\be
\Om_i=\frac{1}{t}\int_o^t dt' \Om_i(t'),
\en
and
\be
\Om_o=\frac{1}{t}\int_o^t dt' \Om_o(t').
\en
The condition for a near r:s resonance leads to an asymptotic condition
for large  $t$, which requires that  
\be
\frac{n_i}{n_o}=\frac{r}{s},
\en
and 
\be
\label{om}
\Om_i=\Om_o=\Om.
\en
Hence, according to Eqs. \ref{wi} and \ref{wo}, 
$\Om=r\om_o-s\om_i)/(r-s)$ which for a 2:1 resonance gives
$\Om=2\om_o-\om_i$ (Lee 2002). 
For negative values of $\Om$, which corresponds to  retrograde motion
for the mean position of the periastron of both planets,
the mean periods $P_i=2\pi/\om_i$, $P_o=2\pi/\om_o$ and $P=2\pi/|\Om|$
satisfy the relation
\be
\frac{P_o}{P_i}=\frac{r}{s}+(\frac{r}{s}-1)\frac{P_o}{P}.
\en
For the case of GJ876, $r=2$ and $s=1$, we obtain Eq. \ref{ratio1}

The angular  oscillations of the  periastron of the inner and outer
planets relative to its mean precession, which have been discussed 
in (Lee 2002;Laughlin 2005), is  given by 
\be
\theta_i=\int_o^t dt'(\Om_i(t')-\Om)
\en 
\be
\theta_o=\int_o^t dt' (\Om_o(t')-\Om)
\en 
Our numerical results are  shown in Figs. 8 and 9, giving  
maximum displacements $|\theta_i|=6^0$ and $|\theta_o|=50^0$ 
which are somewhat different from  those reported in (Laughlin 2005).

\subsection*{Figures}

\begin{figure}
\begin{center}
\epsfxsize=\columnwidth
\epsfig{file=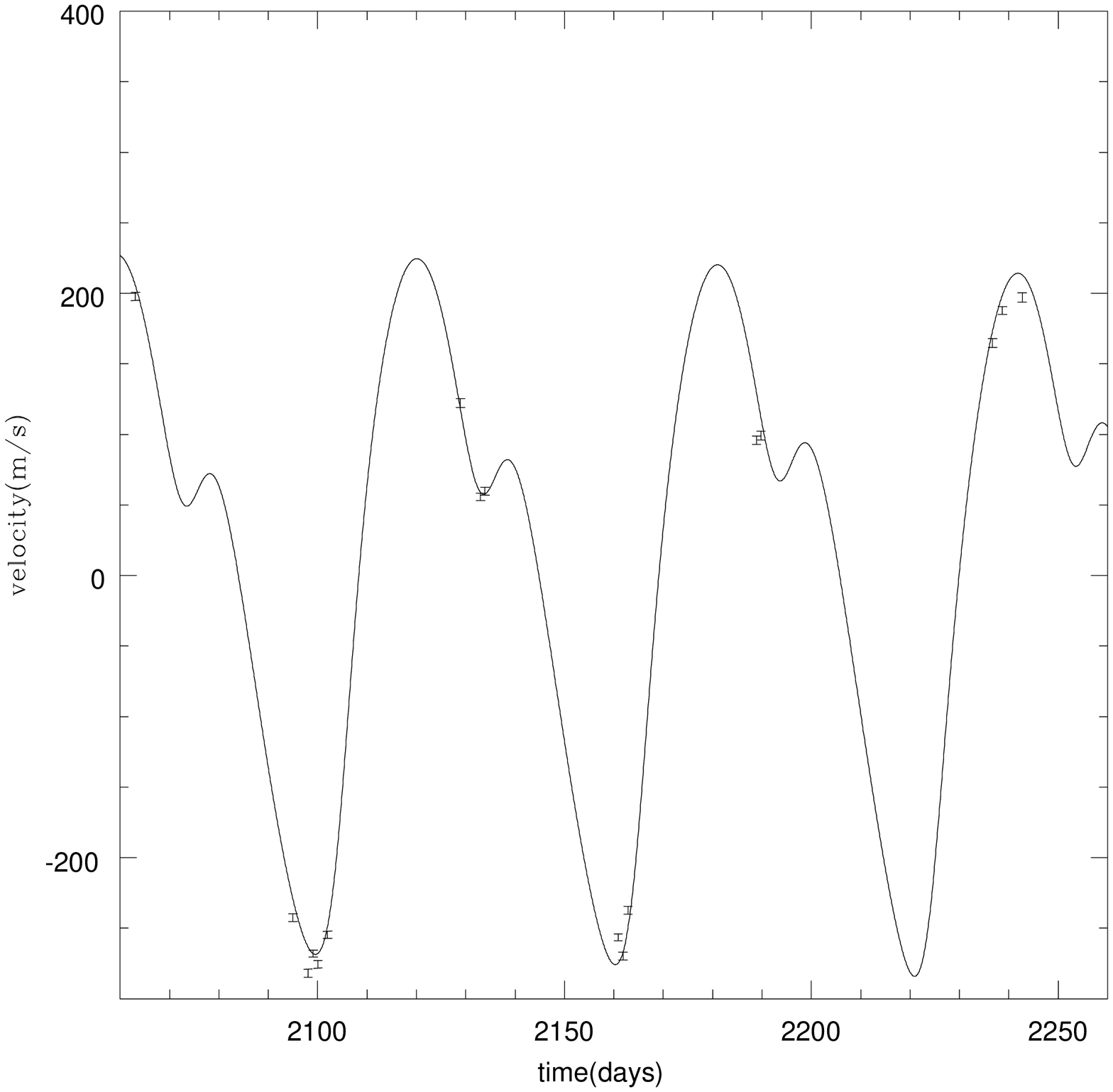, width=15cm}
\end{center}
\caption{ New Keck radial velocities (Laughlin et. al 2005)  
and calculation based on a fit (Nauenberg 2002) 
to the early  Keck data (Marcy et al. 2001)
}
\label{Fig. 1}
\end{figure}

\begin{figure}
\begin{center}
\epsfxsize=\columnwidth
\epsfig{file=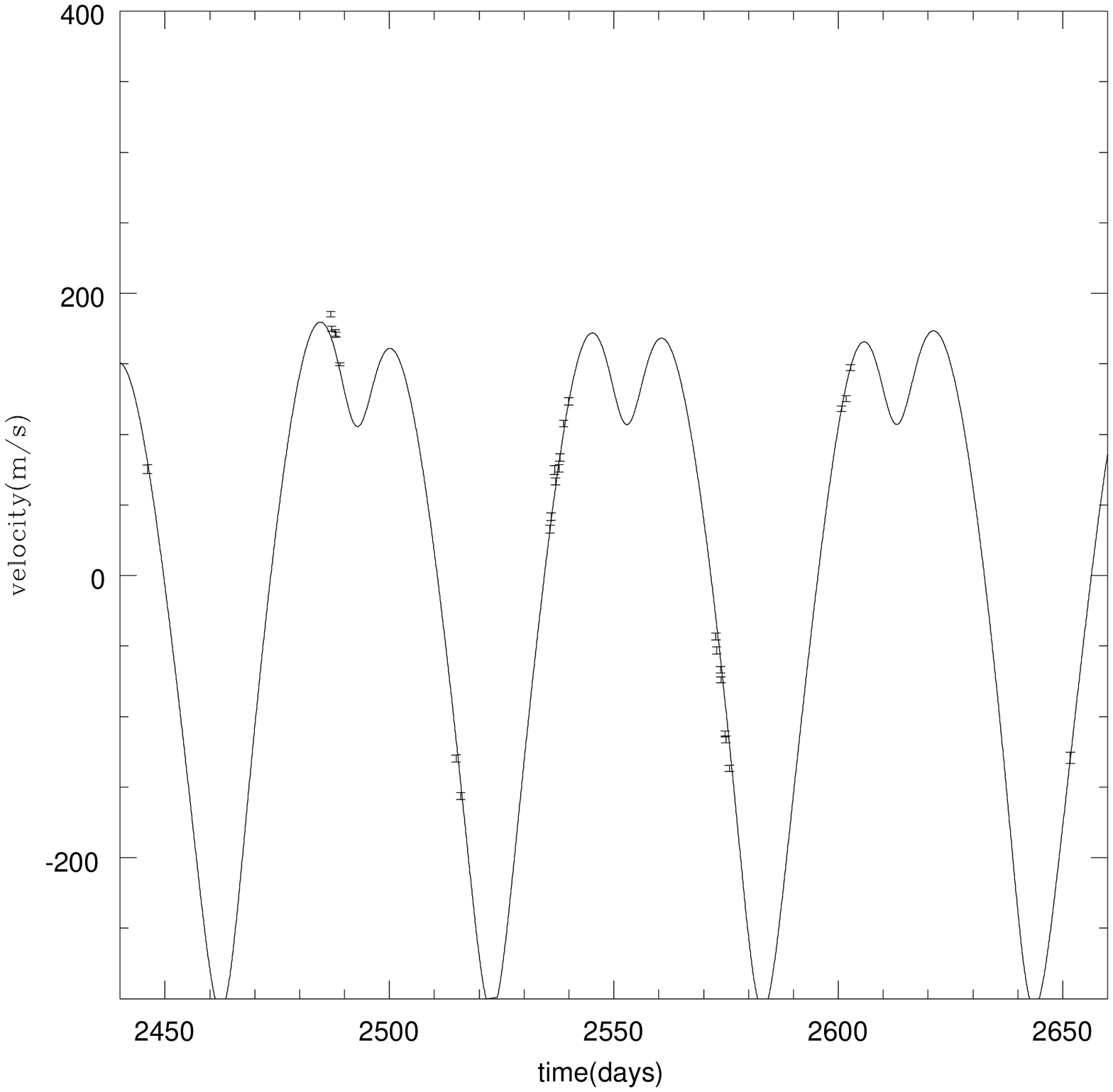, width=15cm}
\end{center}
\caption{ New Keck radial velocities (Laughlin et. al 2005)  
and calculation based on a fit (Nauenberg 2002) 
to the early Keck data of Marcy et al. (2001)
}
\label{Fig. 1}
\end{figure}

\begin{figure}
\begin{center}
\epsfxsize=\columnwidth
\epsfig{file=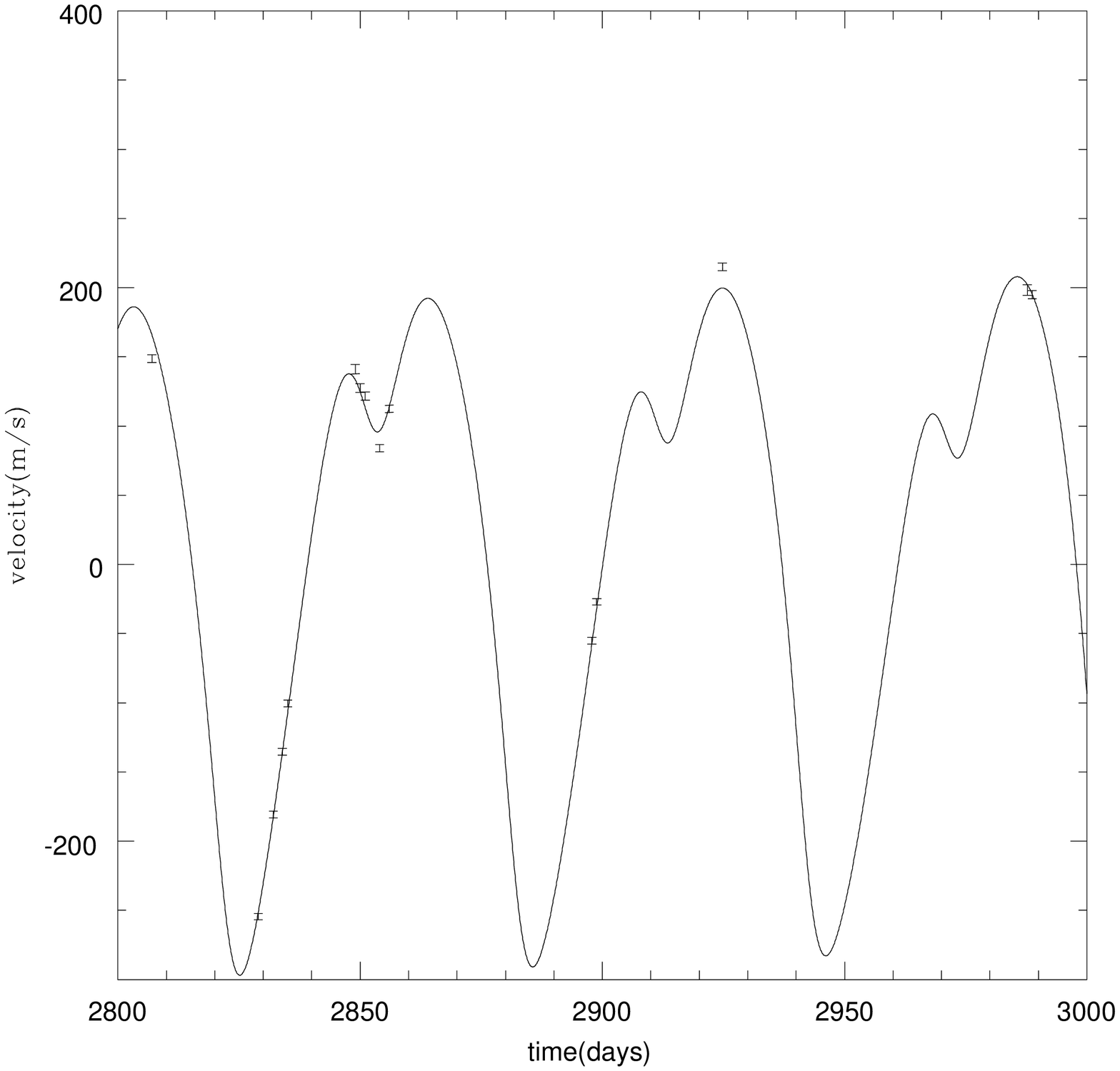, width=15cm}
\end{center}
\caption{ New Keck radial velocities (Laughlin et. al 2005)  
and calculations based on a fit (Nauenberg 2002) 
to the early Keck data Marcy et al. (2001)
}
\label{Fig. 2}
\end{figure}

\begin{figure}
\begin{center}
\epsfxsize=\columnwidth
\epsfig{file=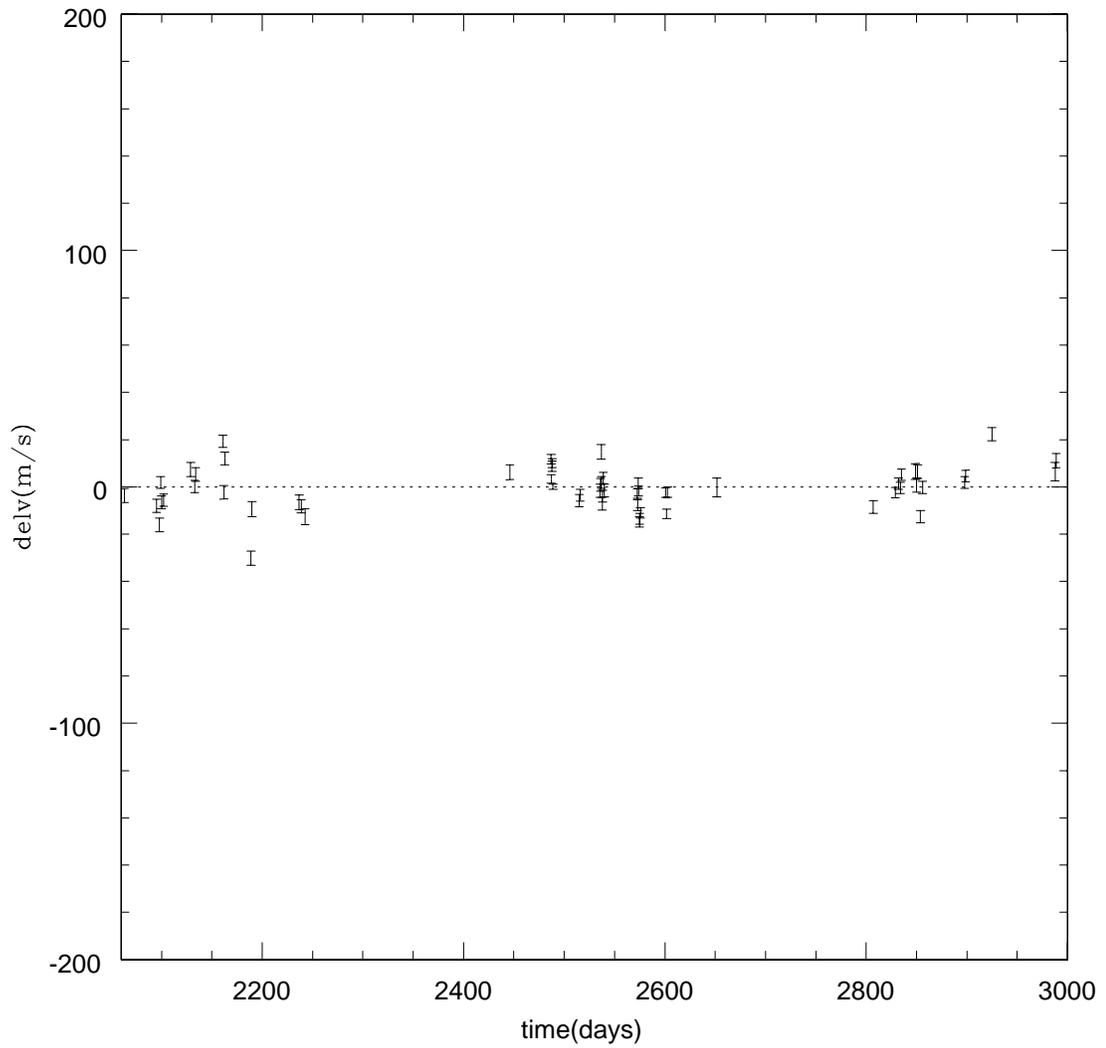, width=15cm}
\end{center}
\caption{Scatter of new Keck radial velocities from 
a fit (Nauenberg 2002) to the early Keck data of Marcy et al.(2001)
}
\label{Fig. 4}
\end{figure}

\begin{figure}
\begin{center}
\epsfxsize=\columnwidth
\epsfig{file=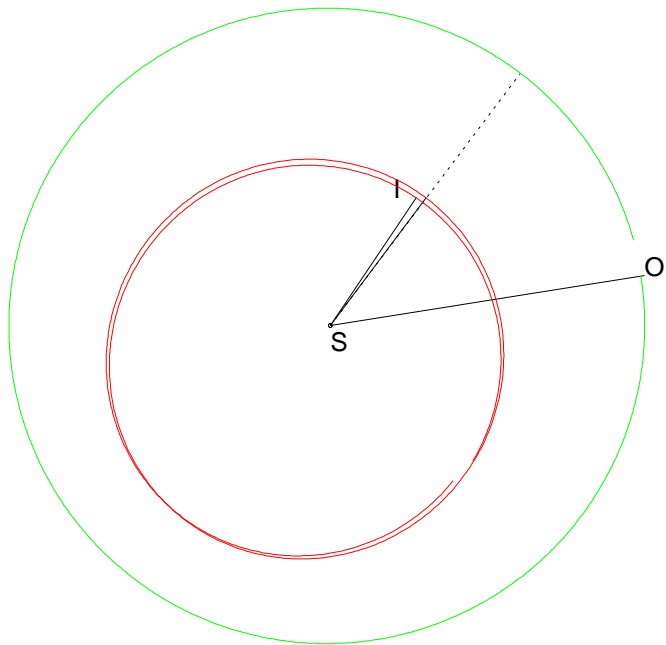, width=15cm}
\end{center}
\caption{  Two consecutive orbits of the
inner planet (red curve), and during a single  orbit
of the outer planet (green curve). The black lines
indicate the  positions of periastron for the
inner planet at $I$, and the outer planet at $O$
}
\label{Fig. 5}
\end{figure}

\begin{figure}
\begin{center}
\epsfxsize=\columnwidth
\epsfig{file=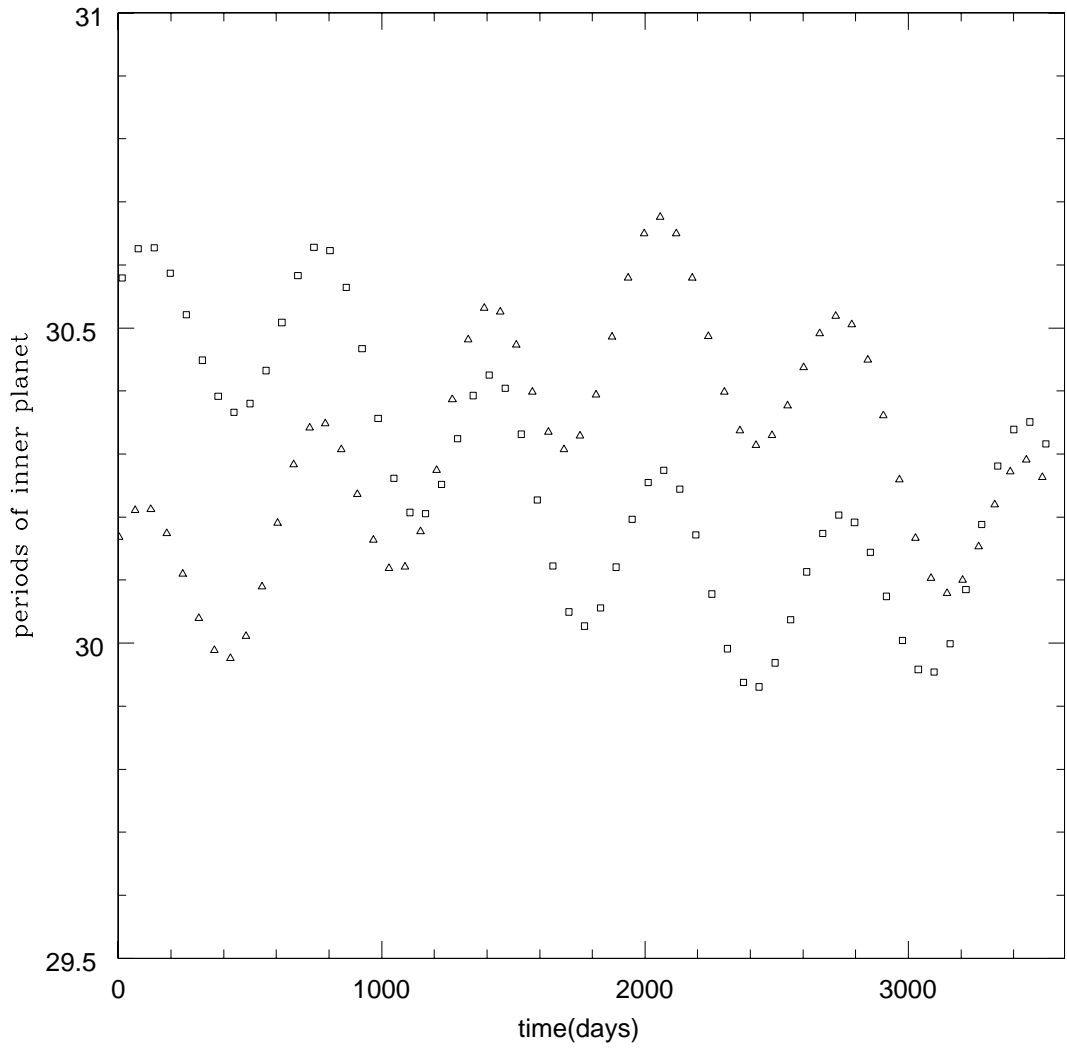, width=15cm}
\end{center}
\caption{ Oscillations of the two periods of the inner planet
in its consecutive orbits, indicated by triangles 
and squares.
}
\label{Fig. 6}
\end{figure}

\begin{figure}
\begin{center}
\epsfxsize=\columnwidth
\epsfig{file=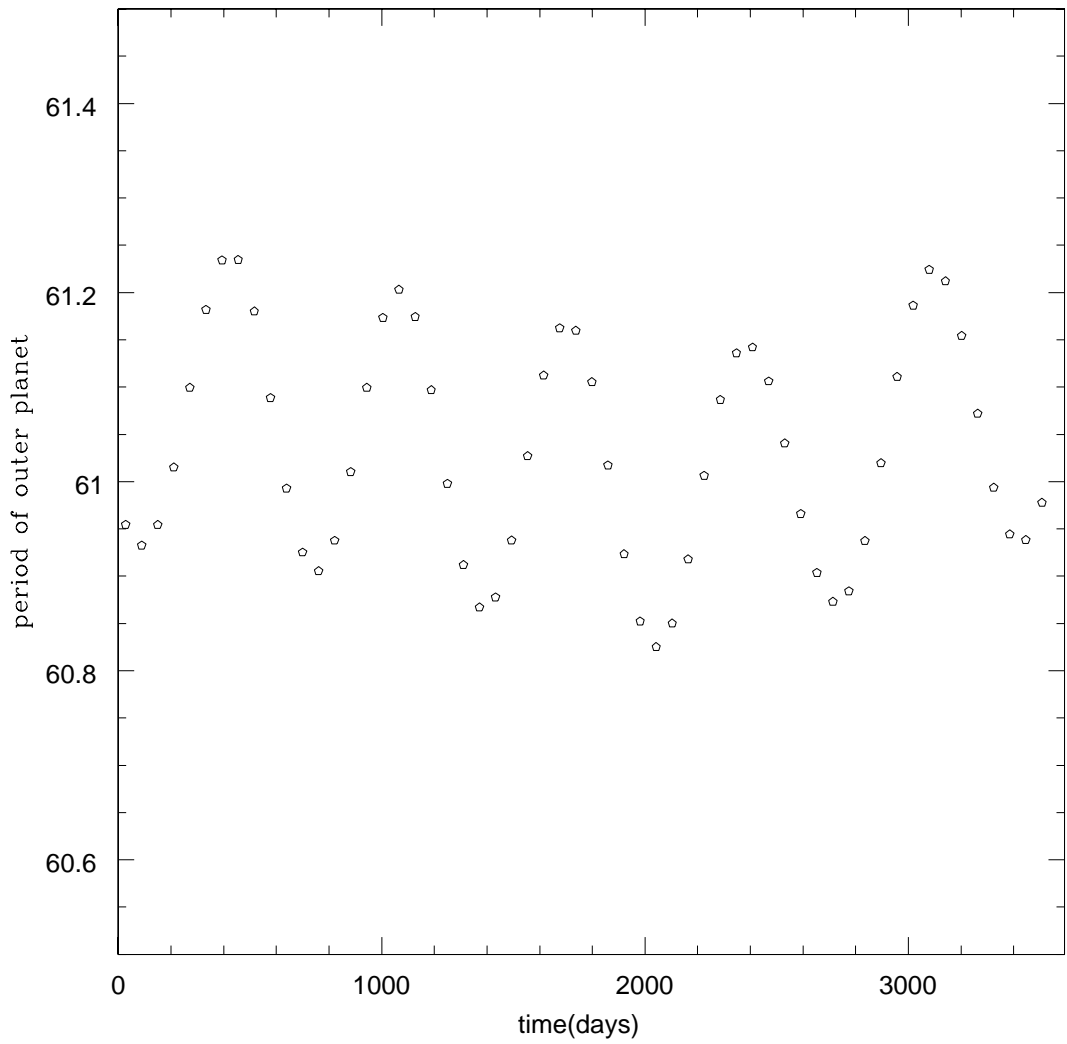, width=15cm}
\end{center}
\caption{Oscillations of the period of the  outer planet
}
\label{Fig. 7}
\end{figure}

\begin{figure}
\begin{center}
\epsfxsize=\columnwidth
\epsfig{file=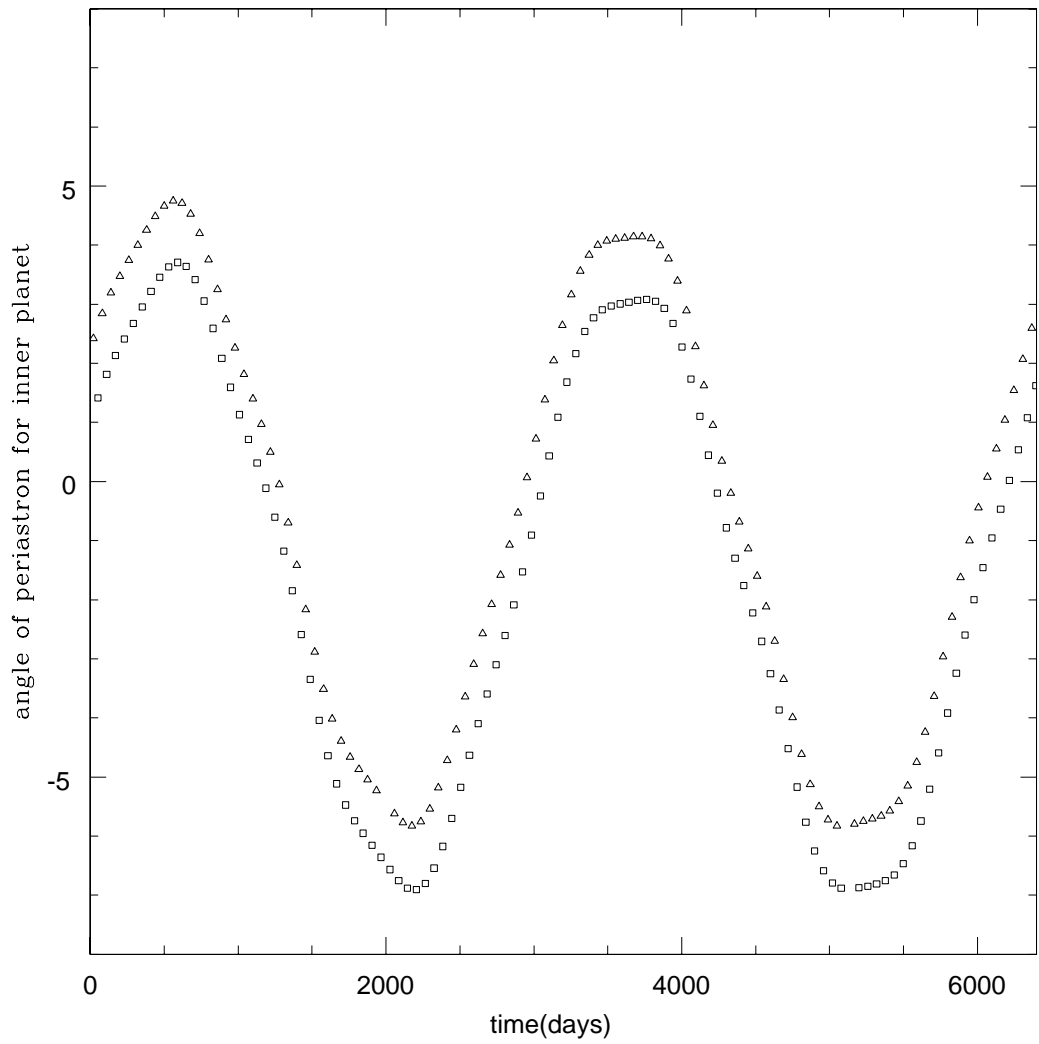, width=15cm}
\end{center}
\caption{ The  periastron angle for 
the two consecutive orbits of the inner planet
shown by triangles and squares
}
\label{Fig. 8}
\end{figure}

\begin{figure}
\begin{center}
\epsfxsize=\columnwidth
\epsfig{file=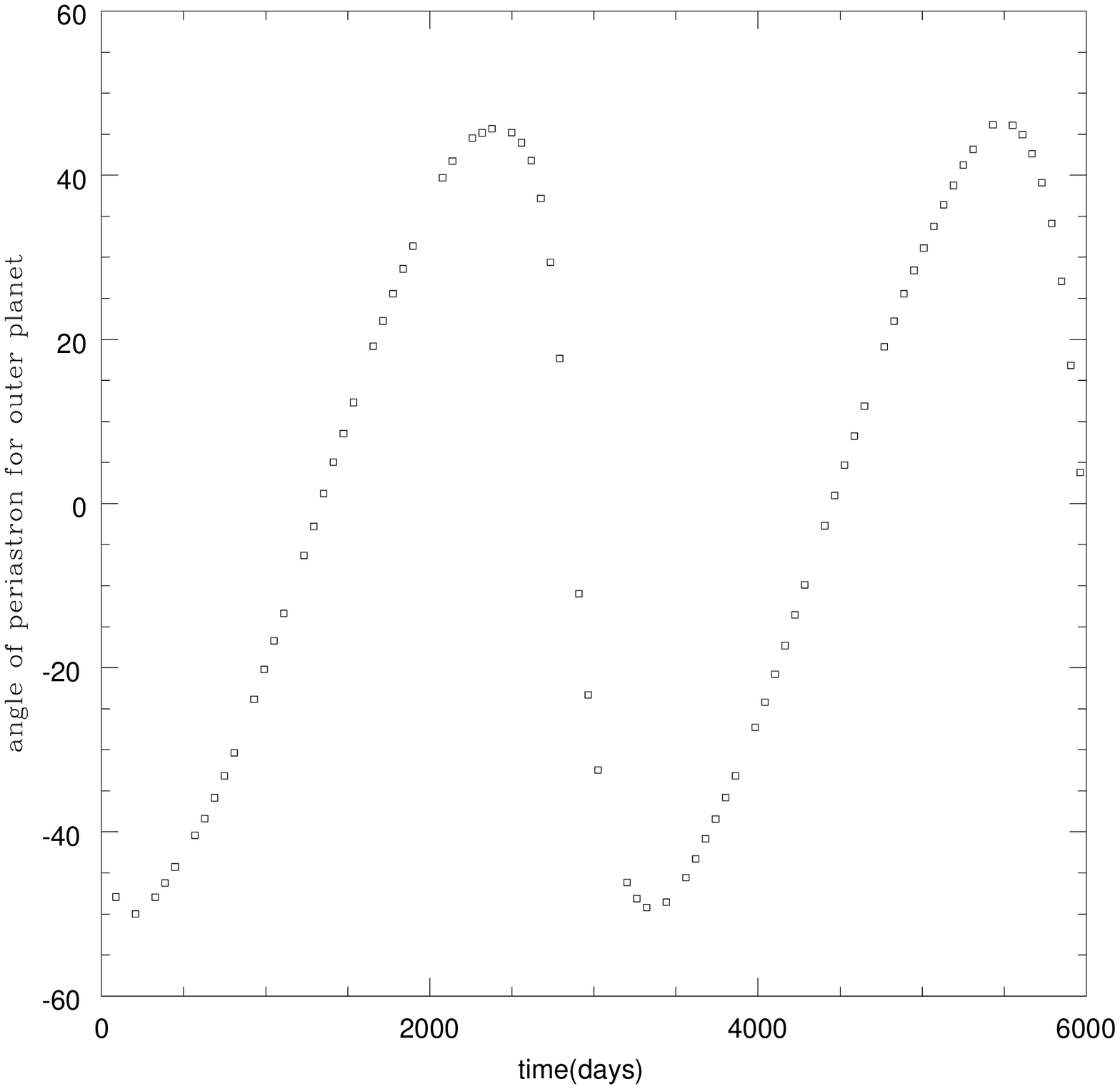, width=15cm}
\end{center}
\caption{ The  periastron angle for
the outer planet
}
\label{Fig. 9}
\end{figure}


\begin{thebibliography}{30}
\bibitem{lau1}  Laughlin, G., Butler, R.P., Fisher, D.A., Marcy, G.W.,
                Vogt, S.S., \&Wolf, A. 2005, AJ, 622, 1182

\bibitem{lau2}  Laughlin, G., \& Chambers, J.E.  2001, ApJ, 567, 594

\bibitem{peale1} Lee, M.H., \& Peale, S.J. 2002, ApJ, 567, 594

\bibitem{marcy1} Marcy, G.W., Butler, R.P., Fischer, D.A., Vogt, S.S.,
                 Lissauer, J.J.,\& Rivera, E.J. 2001, ApJ, 556, 396

\bibitem {mike1} Nauenberg, M. 2002, ApJ, 568, 369

\bibitem {rivera1} Rivera, E.J., \& Lissauer, J.J. 2001, ApJ, 559, 392

\end{thebibliography}
\end{document}